\documentclass{moriond}

\usepackage{amssymb,amsmath}
\usepackage{multirow}
\usepackage{pifont}

\def\be{\begin{equation}}
\def\ee{\end{equation}}
\def\bea{\begin{eqnarray}}
\def\eea{\end{eqnarray}}

\newcommand{\macro}[1]{\textcolor{black}{#1}}

\newcommand{\DISTANCECOMPACTOneFiveZeroNineOneFourCat}{\macro{\ensuremath{440_{-170}^{+150}}}} 
\newcommand{\MTOTSCOMPACTOneFiveZeroNineOneFourCat}{\macro{\ensuremath{66.1_{-3.3}^{+3.8}}}} 
\newcommand{\PEMATCHSNRCOMPACTOneFiveZeroNineOneFourCat}{\macro{\ensuremath{25.3_{-0.2}^{+0.1}}}} 

\newcommand{\DISTANCECOMPACTOneFiveOneZeroOneTwoCat}{\macro{\ensuremath{1080_{-490}^{+550}}}} 
\newcommand{\MTOTSCOMPACTOneFiveOneZeroOneTwoCat}{\macro{\ensuremath{37.2_{-3.9}^{+10.6}}}} 
\newcommand{\PEMATCHSNRCOMPACTOneFiveOneZeroOneTwoCat}{\macro{\ensuremath{9.2_{-0.4}^{+0.3}}}} 

\newcommand{\DISTANCECOMPACTOneFiveOneTwoTwoSixCat}{\macro{\ensuremath{450_{-190}^{+180}}}} 
\newcommand{\MTOTSCOMPACTOneFiveOneTwoTwoSixCat}{\macro{\ensuremath{21.5_{-1.5}^{+6.2}}}} 
\newcommand{\PEMATCHSNRCOMPACTOneFiveOneTwoTwoSixCat}{\macro{\ensuremath{12.4_{-0.3}^{+0.2}}}} 

\newcommand{\DISTANCECOMPACTOneSevenZeroOneZeroFourCat}{\macro{\ensuremath{990_{-430}^{+440}}}} 
\newcommand{\MTOTSCOMPACTOneSevenZeroOneZeroFourCat}{\macro{\ensuremath{51.0_{-4.1}^{+5.3}}}} 
\newcommand{\PEMATCHSNRCOMPACTOneSevenZeroOneZeroFourCat}{\macro{\ensuremath{14.0_{-0.3}^{+0.2}}}} 

\newcommand{\DISTANCECOMPACTOneSevenZeroSixZeroEightCat}{\macro{\ensuremath{320_{-110}^{+120}}}} 
\newcommand{\MTOTSCOMPACTOneSevenZeroSixZeroEightCat}{\macro{\ensuremath{18.6_{-0.7}^{+3.2}}}} 
\newcommand{\PEMATCHSNRCOMPACTOneSevenZeroSixZeroEightCat}{\macro{\ensuremath{15.6_{-0.3}^{+0.2}}}} 

\newcommand{\DISTANCECOMPACTOneSevenZeroSevenTwoNineCat}{\macro{\ensuremath{2840_{-1360}^{+1400}}}} 
\newcommand{\MTOTSCOMPACTOneSevenZeroSevenTwoNineCat}{\macro{\ensuremath{84.4_{-11.1}^{+15.8}}}} 
\newcommand{\PEMATCHSNRCOMPACTOneSevenZeroSevenTwoNineCat}{\macro{\ensuremath{10.8_{-0.5}^{+0.4}}}} 

\newcommand{\DISTANCECOMPACTOneSevenZeroEightZeroNineCat}{\macro{\ensuremath{1030_{-390}^{+320}}}} 
\newcommand{\MTOTSCOMPACTOneSevenZeroEightZeroNineCat}{\macro{\ensuremath{59.0_{-4.1}^{+5.4}}}} 
\newcommand{\PEMATCHSNRCOMPACTOneSevenZeroEightZeroNineCat}{\macro{\ensuremath{12.7_{-0.3}^{+0.2}}}} 

\newcommand{\DISTANCECOMPACTOneSevenZeroEightOneFourCat}{\macro{\ensuremath{600_{-220}^{+150}}}} 
\newcommand{\MTOTSCOMPACTOneSevenZeroEightOneFourCat}{\macro{\ensuremath{55.9_{-2.6}^{+3.4}}}} 
\newcommand{\PEMATCHSNRCOMPACTOneSevenZeroEightOneFourCatLI}{\macro{\ensuremath{17.8_{-0.3}^{+0.3}}}} 

\newcommand{\DISTANCECOMPACTOneSevenZeroEightOneEightCat}{\macro{\ensuremath{1060_{-380}^{+420}}}} 
\newcommand{\MTOTSCOMPACTOneSevenZeroEightOneEightCat}{\macro{\ensuremath{62.2_{-4.1}^{+5.2}}}} 
\newcommand{\PEMATCHSNRCOMPACTOneSevenZeroEightOneEightCat}{\macro{\ensuremath{11.9_{-0.4}^{+0.3}}}} 

\newcommand{\DISTANCECOMPACTOneSevenZeroEightTwoThreeCat}{\macro{\ensuremath{1940_{-900}^{+970}}}} 
\newcommand{\MTOTSCOMPACTOneSevenZeroEightTwoThreeCat}{\macro{\ensuremath{68.7_{-8.1}^{+10.8}}}} 
\newcommand{\PEMATCHSNRCOMPACTOneSevenZeroEightTwoThreeCat}{\macro{\ensuremath{12.0_{-0.3}^{+0.2}}}} 

\newcommand{\MaximumImprovementFromPreviousBoundsOTwoTGR}{\macro{\ensuremath{2.5}}}

\newcommand{\NinetyPercentCombinedGravitonMassBoundScaledOTwoTGR}{4.7}
\newcommand{\NinetyPercentCombinedGravitonMassBoundOTwoTGRImprovementFromGWOneSevenZeroOneZeroFourPaper}{1.6}

\newcommand{\IMRP}{\textsc{IMRPhenomPv2}}
\newcommand{\SEOB}{\textsc{SEOBNRv4}}

\newcommand{\cmark}{\ding{51}}

\newcommand{\rhores}{\mathrm{SNR}_{90}}
\newcommand{\rhonoise}{\rhores^\mathrm{n}}

\newcommand{\Photo}{\includegraphics[height=35mm]{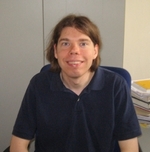}}

\begin{document}
\vspace*{4cm}
\title{Summary of Tests of General Relativity with the Binary Black Hole Signals from the LIGO-Virgo Catalog GWTC-1}

\author{N.~K.~Johnson-McDaniel \\ (for the LIGO Scientific Collaboration and Virgo Collaboration)}

\address{Department of Applied Mathematics and Theoretical Physics, Centre for Mathematical Sciences, University of Cambridge,  Cambridge,  CB3 0WA, United Kingdom}

\maketitle\abstracts{
We summarize the four tests of general relativity carried out using the ten binary black hole signals detected by Advanced LIGO and Advanced Virgo and included in their first catalog, GWTC-1. These events provide unprecedented opportunities for testing the predictions of general relativity for the gravitational waveforms from these highly dynamical, strong-field events. The first two tests check the consistency of the residuals with noise and the consistency of the low- and high-frequency parts of the signal. The other two tests check that parameterized deviations in the waveform model---including in the post-Newtonian coefficients---are consistent with zero, and that the propagation of the waves is nondispersive. These tests reveal no evidence for deviations from general relativity, and the combined constraints improve previous results by factors of up to $\MaximumImprovementFromPreviousBoundsOTwoTGR$. We also check that the binary black hole signals observed by all three detectors do not give stronger constraints on alternative polarizations than those obtained from GW170817.
}

\section{Introduction}

General relativity (GR) has been subjected to a wide variety of tests in its over 100 years of existence~\cite{lrr-2014-4}. However, it is only in the past few years
that we are able to test it in some of the most extreme regimes in the Universe, the coalescences of compact binaries. This is possible through the detection of gravitational waves from these systems by the Advanced LIGO~\cite{TheLIGOScientific:2014jea} and Advanced Virgo~\cite{TheVirgo:2014hva} detectors. In the first two observing runs of the advanced detector era, LIGO and Virgo detected significant signals from ten binary black hole coalescences and one binary neutron star coalescence, given in their first catalog, GWTC-1~\cite{O2:Catalog,GWOSC:GWTC}. Here we will discuss the tests of general relativity performed using those ten binary black hole coalescences. These results are given in a detailed paper~\cite{LIGOScientific:2019fpa} (later referred to as ``the main paper''), which we summarize here. The tests of general relativity performed by the LIGO and Virgo collaborations using the binary neutron star coalescence GW1710817 and its electromagnetic counterparts are described in other papers~\cite{Monitor:2017mdv,bns-tgr}. 

The binary black hole events provide very clean laboratories with which to test general relativity in the highly dynamical, strong-field regime. Indeed, observations of compact binary coalescences are at present our only way of probing gravity in such regimes. These binaries reach speeds of roughly $0.5c$ close to merger (see, e.g., Fig.~2 from the GW150914 detection paper~\cite{GW150914_paper}) and have spacetime curvature scales at least as small as $\sim 14$~km.\footnote{This curvature scale is obtained from the inverse fourth root of the Kretschmann scalar at the horizon of the individual black hole with the lowest $90\%$ credible level upper bound on its mass ($8.9M_\odot$, for the lighter black hole in the source of GW170608). We use the Schwarzschild value to be conservative---spinning black holes have more strongly curved spacetime for a given total mass, but only by at most a factor of $\sim 3$ by this measure. See, e.g., Eq.~(5.47) in Poisson's textbook~\cite{2004rtmb.book.....P} for the expression for the Kretschmann scalar for Kerr black holes.} These binary black holes are also at very large distances compared to their orbital separations (factors of $\gtrsim 10^{20}$, considering the evolution since the binary enters the LIGO/Virgo band), which makes these signals an excellent place to test propagation effects like dispersion.

In the absence of detailed calculations of gravitational waveforms from binary black holes in alternative theories of gravity (that give different predictions from GR), we perform null tests, either looking for residual signals after subtracting the best-fit GR waveform, checking for consistency of different parts of the signal, or looking for deviations from the GR predictions for waveform coefficients or in the GR predictions for nondispersive propagation of gravitational waves.

\section{Events}
\label{sec:events}

\begin{table*}
\caption{\label{tab:events}
The GW events considered in this paper, separated by observing run. The first block of columns gives the names of the events and lists some of their relevant
properties obtained using GR waveforms (luminosity distance $D_\text{L}$ and source frame total mass $M_\text{tot}$). The middle column gives the signal-to-noise ratio (SNR) from the matched filter analysis with GR waveforms. The last block of columns indicates which GR tests are performed on a given event: RT = residuals test (Sec.~\ref{sec:residuals}); IMR = inspiral-merger-ringdown consistency test (Sec.~\ref{sec:imr-test}); PI \& PPI = parameterized tests of GW generation for inspiral and post-inspiral phases (Sec.~\ref{sec:gwnature}); MDR = modified GW dispersion relation (Sec.~\ref{sec:propagation}).
The events with bold names are used to obtain the combined results for each test. Adapted from Table~I in the main paper.}
\begin{tabular}{| c c c c c c c c c c c |}
\hline
\multirow{2}{*}{Event} & \multicolumn{2}{c}{Properties} & \hphantom{X} & \multirow{2}{*}{SNR} & \hphantom{X} & \multicolumn{5}{c|}{GR tests performed}\\
\cline{2-3}
\cline{7-11}
& $D_\text{L}$ & $M_\text{tot}$ & & & & RT & IMR & PI \ & PPI \ & MDR\\
& [Mpc] & [$M_\odot$] & & & & & & & &\\
\hline
\textbf{GW150914} & \DISTANCECOMPACTOneFiveZeroNineOneFourCat & \MTOTSCOMPACTOneFiveZeroNineOneFourCat & & \PEMATCHSNRCOMPACTOneFiveZeroNineOneFourCat & & \cmark & \cmark & \cmark & \cmark & \cmark\\[0.075cm]
GW151012 & \DISTANCECOMPACTOneFiveOneZeroOneTwoCat & \MTOTSCOMPACTOneFiveOneZeroOneTwoCat & & \PEMATCHSNRCOMPACTOneFiveOneZeroOneTwoCat & & \cmark & -- & -- & \cmark & \cmark\\[0.075cm]
\textbf{GW151226} &\DISTANCECOMPACTOneFiveOneTwoTwoSixCat & \MTOTSCOMPACTOneFiveOneTwoTwoSixCat & & \PEMATCHSNRCOMPACTOneFiveOneTwoTwoSixCat & & \cmark & -- & \cmark & -- & \cmark\\[0.075cm]
\hline
\textbf{GW170104} &\DISTANCECOMPACTOneSevenZeroOneZeroFourCat & \MTOTSCOMPACTOneSevenZeroOneZeroFourCat & & \PEMATCHSNRCOMPACTOneSevenZeroOneZeroFourCat & & \cmark & \cmark & \cmark & \cmark & \cmark\\[0.075cm]
\textbf{GW170608} &  \DISTANCECOMPACTOneSevenZeroSixZeroEightCat & \MTOTSCOMPACTOneSevenZeroSixZeroEightCat & & \PEMATCHSNRCOMPACTOneSevenZeroSixZeroEightCat & & \cmark & -- & \cmark & \cmark & \cmark\\[0.075cm]
GW170729 & \DISTANCECOMPACTOneSevenZeroSevenTwoNineCat & \MTOTSCOMPACTOneSevenZeroSevenTwoNineCat & & \PEMATCHSNRCOMPACTOneSevenZeroSevenTwoNineCat & & \cmark & \cmark & -- & \cmark & \cmark\\[0.075cm]
\textbf{GW170809} & \DISTANCECOMPACTOneSevenZeroEightZeroNineCat & \MTOTSCOMPACTOneSevenZeroEightZeroNineCat & & \PEMATCHSNRCOMPACTOneSevenZeroEightZeroNineCat & & \cmark & \cmark & -- & \cmark & \cmark\\[0.075cm]
\textbf{GW170814} & \DISTANCECOMPACTOneSevenZeroEightOneFourCat & \MTOTSCOMPACTOneSevenZeroEightOneFourCat & & \PEMATCHSNRCOMPACTOneSevenZeroEightOneFourCatLI & & \cmark & \cmark & \cmark & \cmark & \cmark\\[0.075cm]
GW170818 & \DISTANCECOMPACTOneSevenZeroEightOneEightCat & \MTOTSCOMPACTOneSevenZeroEightOneEightCat & & \PEMATCHSNRCOMPACTOneSevenZeroEightOneEightCat & & \cmark & \cmark & -- & \cmark & \cmark\\[0.075cm]
\textbf{GW170823} & \DISTANCECOMPACTOneSevenZeroEightTwoThreeCat & \MTOTSCOMPACTOneSevenZeroEightTwoThreeCat & & \PEMATCHSNRCOMPACTOneSevenZeroEightTwoThreeCat & & \cmark & \cmark & -- & \cmark & \cmark\\
\hline
\end{tabular}
\end{table*}

We list the events we consider in Table~\ref{tab:events}. In order to check agreement with general relativity for events with somewhat low significance (to see if this can explain why they are less significant than other events), we set a relatively low threshold of a false-alarm rate (FAR) of less than one per year in any of the three detection pipelines considered in the GWTC-1 paper~\cite{O2:Catalog}: two modeled pipelines~\cite{pycbc-github,Canton:2014ena,Usman:2015kfa,Sachdev:2019vvd,Messick:2016aqy}, which use GR waveforms, and one weakly modeled pipeline~\cite{Klimenko:2008fu,Klimenko:2015ypf,TheLIGOScientific:2016uux}, which assumes a chirping signal but no specific GR predictions. The weakly modeled pipeline would likely pick up a sufficiently strong signal that differed from GR predictions, even if it was downweighted by the GR signal consistency tests in the modeled pipelines. However, to obtain our combined results, we only consider very confident events, with FARs less than one per thousand years in both modeled pipelines.

Additionally, we are not able to perform every test on each event: While the residuals and propagation tests are applicable to every event, the consistency test of the low- and high-frequency parts of the signal requires there to be sufficient signal-to-noise ratio (SNR) in both parts of the signal. For the tests of waveform coefficients, one again needs sufficient SNR in either the low- or high-frequency part of the signal to check the coefficients describing those regimes. We chose an SNR of 6 in either the low- or high-frequency part of the signal as the threshold (defining low- and high-frequency appropriately for each test, as discussed in the main paper), and show the events for which one can apply these tests in Table~\ref{tab:events}.

When combining together these results, we follow the spirit of our null test and combine the results assuming that GR is correct (e.g., multiplying together posterior probability distributions from the various events for the tests that produce these). As discussed in the main paper and elsewhere~\cite{Zimmerman:2019wzo}, this is a strong assumption, except in the case of the propagation constraints, where
the dependence on the distance to the source is accounted for in the application of the test. However, as shown in~\cite{Ghosh:2017gfp} this way of combining together results can detect some deviations from GR that are not the same for every signal. There are also recent proposals for how to combine together results without such assumptions~\cite{Isi:2019asy} that may be used in the future.

\section{Waveforms and inference}

All of the tests of GR we consider rely on having accurate predictions for the gravitational waveforms from binary black holes in GR. One needs full numerical simulations to accurately describe the waveforms from the late inspiral and merger of these systems~\cite{Sperhake:2014wpa}. However, such simulations are very computationally expensive, and cannot be run for long enough to describe the full signal in the LIGO/Virgo band except for sufficiently massive systems. One thus constructs fast-to-evaluate model waveforms that use perturbation theory results (notably post-Newtonian theory to describe the inspiral~\cite{Blanchet:2013haa}) and are calibrated to the results of numerical simulations.

We use the \IMRP{} natively frequency domain precessing waveform model~\cite{Hannam:2013oca,Khan:2015jqa} for the primary results in this paper and use the \SEOB{} aligned-spin model~\cite{Bohe:2016gbl} to perform checks of waveform systematics (this is a reduced order model of the natively time-domain effective-one-body model). While there is a precessing effective-one-body waveform model
used to estimate the parameters of the GWTC-1 binary black hole systems~\cite{O2:Catalog}, it does not yet have a reduced order model, making it too slow to be used for most  of the tests we consider. While there is physics missing from these waveform models (notably double-spin precession, higher modes, and eccentricity), we see no evidence in these tests that this missing physics is needed to describe the signals we observe to the accuracy allowed by the noise. As discussed in the main paper, waveform models including these additional effects are now available or being developed
and will be used in future applications of these tests of GR.

We estimate the parameters describing the binaries and any additional non-GR parameters in the framework of Bayesian inference using the \textsc{LALInference} 
code~\cite{Veitch:2014wba} in the LIGO Scientific Collaboration Algorithm Library Suite (LALSuite)~\cite{lalsuite}. We estimate the noise power spectral density using the \textsc{BayesWave} code~\cite{Cornish:2014kda,Littenberg:2014oda}, as described in Appendix~B of~\cite{O2:Catalog}.

\section{Residuals test}
\label{sec:residuals}

\begin{figure} 
	\centering
	\includegraphics[width=0.45\columnwidth]{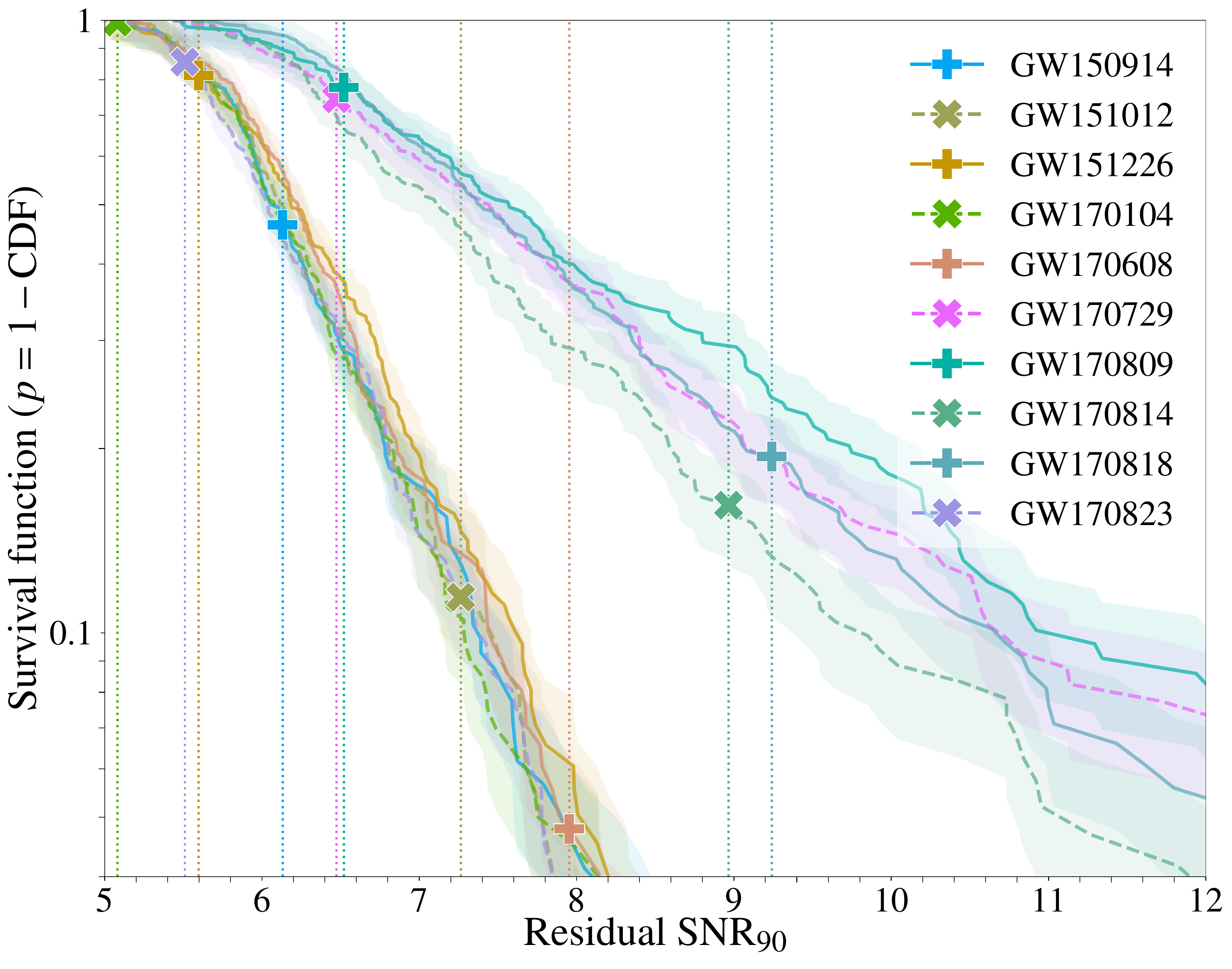}
	\qquad
	\includegraphics[width=0.375\columnwidth]{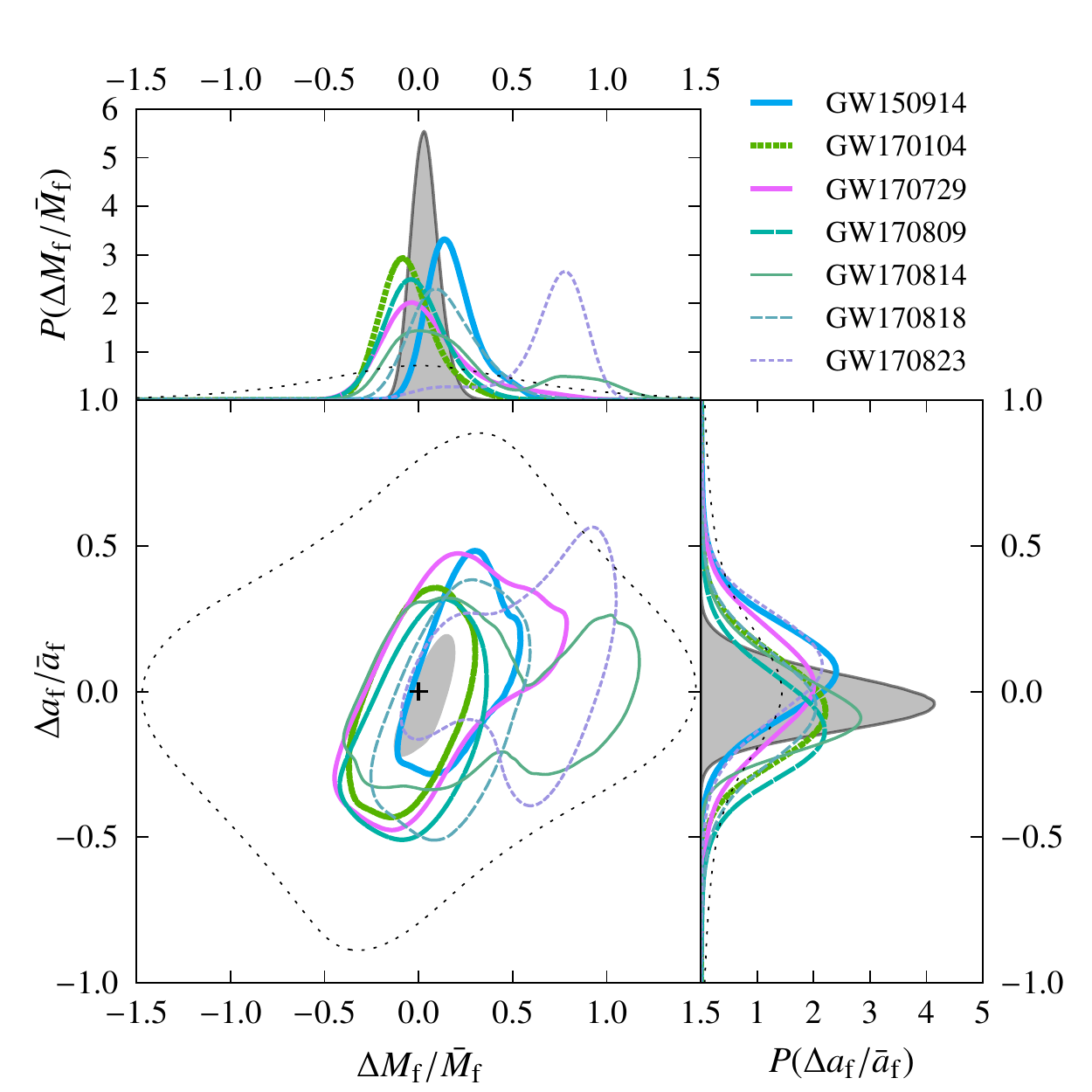}
	\caption{\emph{(Left)} Survival function ($p = 1-\mathrm{CDF}$) of the 90\%-credible upper limit on the noise-only network SNR ($\rhores$) for each event ({solid or dashed} curves), compared to the measured residual values (vertical dotted lines).
For each event, the value of the survival function at the measured $\rhores$ gives the $p$-value (markers). 
The colored bands correspond to uncertainty regions for a Poisson process and have half width $\pm p/\sqrt{N}$, with $N$ the number of noise-only instantiations that yielded $\rhonoise$ greater than the abscissa value. \emph{(Right)} Results of the inspiral-merger-ringdown consistency test for the selected BBH events (see Table~\ref{tab:events}). The main panel shows 90\% credible regions of the posterior distributions of $(\Delta M_\mathrm{f}/\bar{M}_\mathrm{f}, \Delta a_\mathrm{f}/\bar{a}_\mathrm{f})$, with the cross marking the expected value for GR. The side panels show the marginalized posteriors for $\Delta M_\mathrm{f}/\bar{M}_\mathrm{f}$ and $\Delta a_\mathrm{f}/\bar{a}_\mathrm{f}$. The thin black dashed curve represents the prior distribution, and the grey shaded areas correspond to the combined posteriors from the five most significant events (as outlined in Sec.~\ref{sec:events} and Table~\ref{tab:events}). Both figures are reproduced from the main paper.}
	\label{fig:residuals_snr_and_imr}
\end{figure}

The first test we consider checks whether the residual remaining after subtracting the maximum likelihood GR waveform (from the analysis in~\cite{O2:Catalog}) is consistent with detector noise. We do this by calculating the SNR of the residuals using the \textsc{BayesWave} code, which describes the signal in the detectors as a sum of incoherent Gaussian noise and an elliptically-polarized coherent signal. We then compare the 90\% credible upper limit on the SNR from \textsc{BayesWave} to the distribution of SNRs obtained from applying the same analysis to 200 different sets of noise-only detector data near each event. This distribution lets us compute a $p$-value from the residuals SNR. The results are shown in Fig.~\ref{fig:residuals_snr_and_imr}. We find that the $p$-values are at least $0.05$ (this smallest value is for GW170608), and the meta $p$-value (checking that the individual $p$-values are uniformly distributed) computed using Fisher's method~\cite{Fisher1948} is $0.4$. Thus, there is no statistically significant evidence for deviations from GR.

\section{Inspiral-merger-ringdown consistency test}
\label{sec:imr-test}

The inspiral-merger-ringdown consistency test~\cite{Ghosh:2016qgn,Ghosh:2017gfp} checks if the final mass and spin inferred from the low- and high-frequency parts of the signal are consistent. By the stationary phase approximation, the low- and high-frequency parts of the signal roughly correspond to the inspiral and post-inspiral (merger-ringdown) parts of the signal, hence the name. The final mass and spin are used for this test because they are generally well-determined quantities, particularly from the merger-ringdown part. The split between the low- and high-frequency parts of the signal is given by the dominant gravitational wave frequency associated with the innermost stable circular orbit of the final black hole (i.e., twice the orbital frequency), using the analysis on the full signal from~\cite{O2:Catalog}.

The analysis applies the same waveform models to the low- and high-frequency parts of the signal, just changing the frequencies over which the likelihood integral is computed, and uses the same fits to numerical relativity simulations to obtain the final mass and spin from the initial masses and spins that parameterize the waveform model. The differences between the two estimates of the same quantities are normalized by the average of the estimates to give scaled quantities that are used when combining together results: $\Delta M_\mathrm{f}/\bar{M}_\mathrm{f} := 2\, (M_\mathrm{f}^{\text{insp}} - M_\mathrm{f}^{\text{post-insp}})/(M_\mathrm{f}^{\text{insp}} + M_\mathrm{f}^{\text{post-insp}})$ and $\Delta a_\mathrm{f}/\bar{a}_\mathrm{f} := 2\, (a_\mathrm{f}^{\text{insp}} - a_\mathrm{f}^{\text{post-insp}})/(a_\mathrm{f}^{\text{insp}} + a_\mathrm{f}^{\text{post-insp}})$. The results are presented in Fig.~\ref{fig:residuals_snr_and_imr}, and all events are consistent with GR, with the GR value being recovered at the $\leq 80\%$ credible level---the largest value is for GW170823. The analysis with \SEOB{} produces similar results.

\section{Parameterized test of gravitational wave generation}
\label{sec:gwnature}

\begin{figure}[tb]
\includegraphics[width=0.4\textwidth]{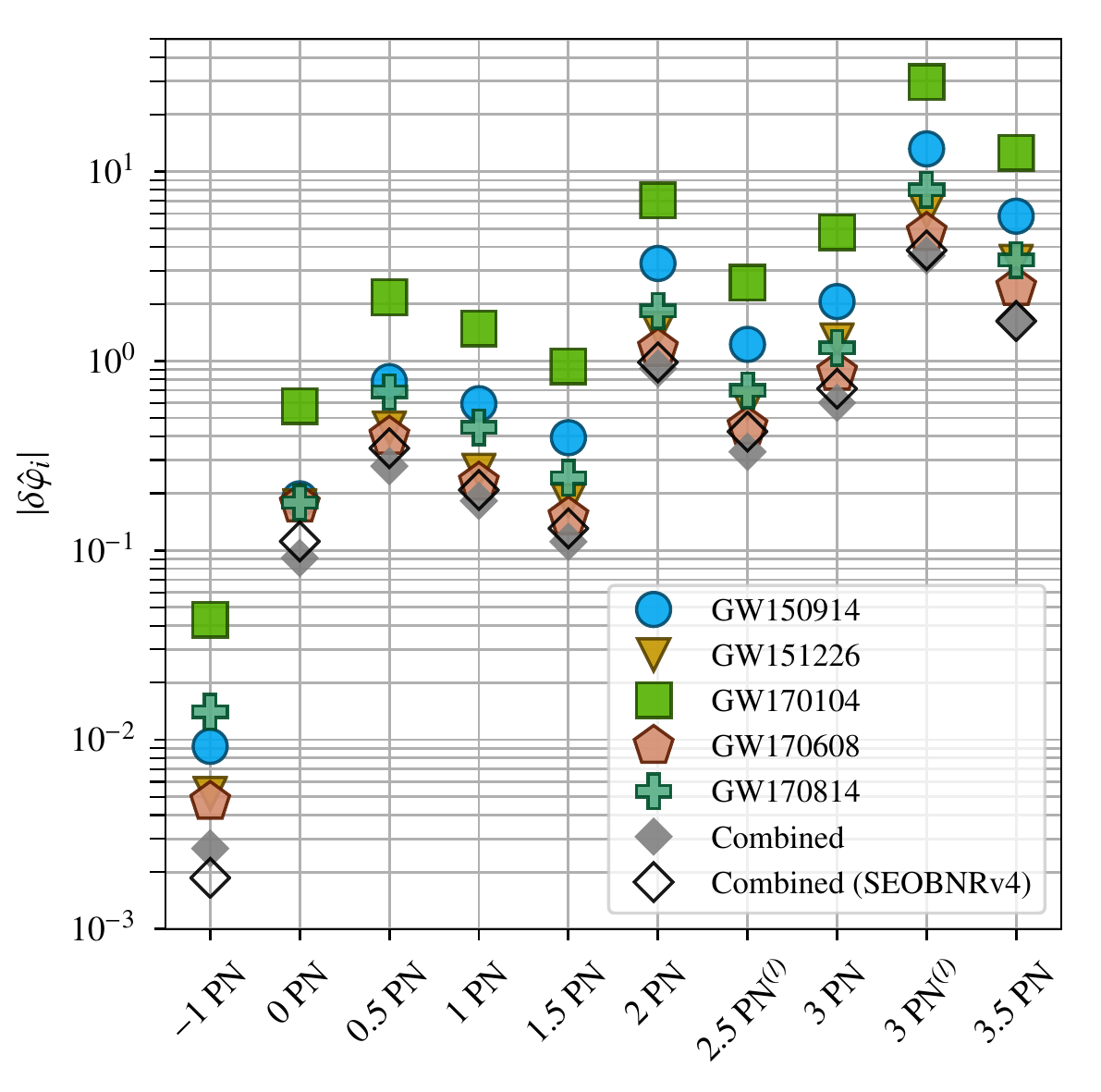}
\qquad
\includegraphics[trim=0 150 0 120,clip,width=0.5\textwidth]{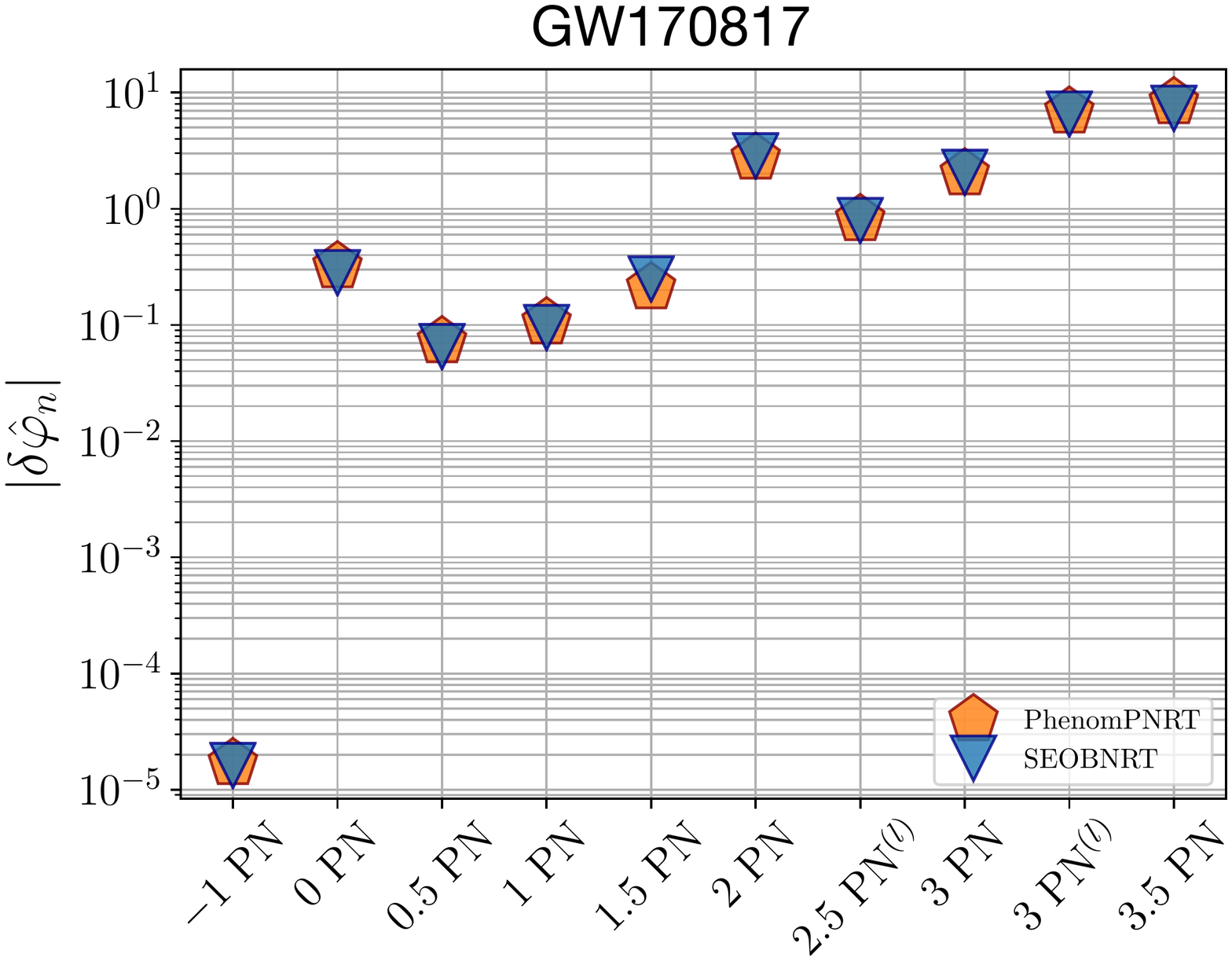}
\\
\includegraphics[width=\textwidth]{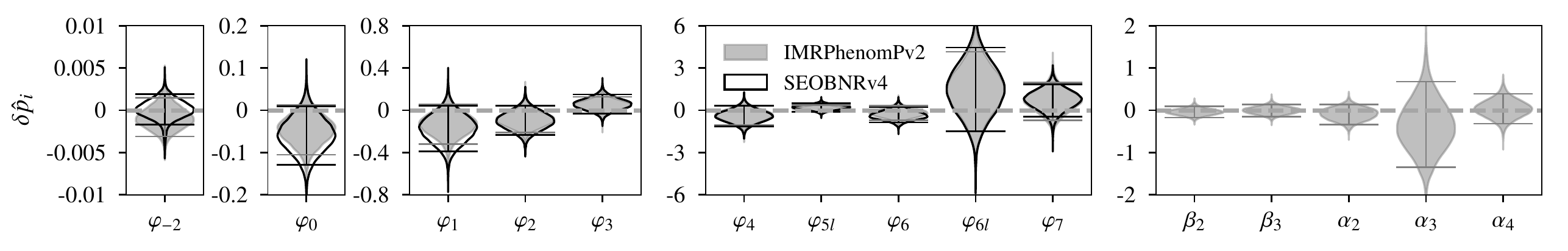}
\caption{\emph{(Top left)} 90\% upper bounds on the absolute magnitude of the GR-violating parameters $\delta\hat\varphi_n$, from $-1$PN through 3.5PN in the inspiral phase.
  At each PN order, we show results obtained from each of the events listed in Table~\ref{tab:events} that cross the SNR threshold in the inspiral regime, analyzed with {\IMRP}.
  Bounds obtained from combining posteriors of events detected with a significance that exceeds a threshold of $\mathrm{FAR < (1000~yr)}^{-1}$ in both modelled searches are shown for both analyses, using {\IMRP} (filled diamonds) and {\SEOB} (empty diamonds). \emph{(Top right)} 90\% upper bounds on deviations $|\delta\hat\varphi_n|$ in the PN coefficients obtained from GW170817 using the extensions of the \IMRP{} and \SEOB{} models to include tidal effects. \emph{(Bottom)} Combined posteriors for parametrized violations of GR, obtained from all events in Table~\ref{tab:events} with a significance of $\mathrm{FAR < (1000~yr)}^{-1}$ in both modeled searches.
The horizontal lines indicate the 90\% credible intervals, and the dashed horizontal line at zero corresponds to the expected GR values.
  The combined posteriors on $\varphi_i$ in the inspiral regime are obtained from the events which in addition exceed the SNR threshold in the inspiral regime (GW150914, GW151226, GW170104, GW170608, and GW170814), analyzed with {\IMRP} (grey shaded region) and {\SEOB} (black outline).
  The combined posteriors on the intermediate and merger-ringdown parameters $\beta_i$ and $\alpha_i$ are obtained from events which exceed the SNR threshold in the post-inspiral regime (GW150914, GW170104, GW170608, GW170809, GW170814, and GW170823), analyzed with {\IMRP}. Figures are reproduced from the main paper and the GW170817 testing GR paper.}
\label{fig:pn_bounds_and_violins}
\end{figure}

This test considers parameterized deviations of various coefficients describing the (frequency-domain) phase of the aligned-spin \textsc{IMRPhenomD} waveform model that underlies \IMRP{}, and checks that these deviations are consistent with zero~\cite{Meidam:2017dgf}. In particular, it considers deviations in all the post-Newtonian (PN) coefficients $\varphi_i$ present in the waveform model [through 3.5PN, or $O(v^7)$, where $v$ is the binary's orbital speed], as well as the possibility for terms at $-1$PN and $0.5$PN, which would be associated with dipole radiation. The test also considers deviations in the phenomenological coefficients $\beta_i$ and $\alpha_i$ describing the intermediate frequency and merger-ringdown portions of the model, respectively. The deviation parameters are denoted by $\delta\hat{p}_i$, where $p\in\{\varphi, \beta, \alpha\}$ and are fractional deviations in all cases except for the $-1$PN and $0.5$PN coefficients, which are zero in GR.

The results are shown in Fig.~\ref{fig:pn_bounds_and_violins}, along with the bounds on deviations in the PN coefficients using GW170817~\cite{bns-tgr}, for comparison. The GW170817 results are obtained using tidal extensions~\cite{Dietrich:2017aum,Dietrich:2018uni} to \IMRP{} and \SEOB{}. The \SEOB{} constraints on the PN coefficients are obtained by tapering the deviation above the end of the inspiral in the \textsc{IMRPhenomD} model, while the \IMRP{} results are obtained by letting the change to the PN coefficient affect the rest of the waveform by the $C^1$ matching used to construct the waveform model. These results thus check for possible systematics in the way of adding the deviations as well as in the waveform modeling. The results are all consistent with GR, and the combined results improve previous results by factors between 1.1 and 1.8.

\section{Parameterized test of gravitational wave propagation}
\label{sec:propagation}

\begin{figure}[tb]
\includegraphics[width=0.48\textwidth]{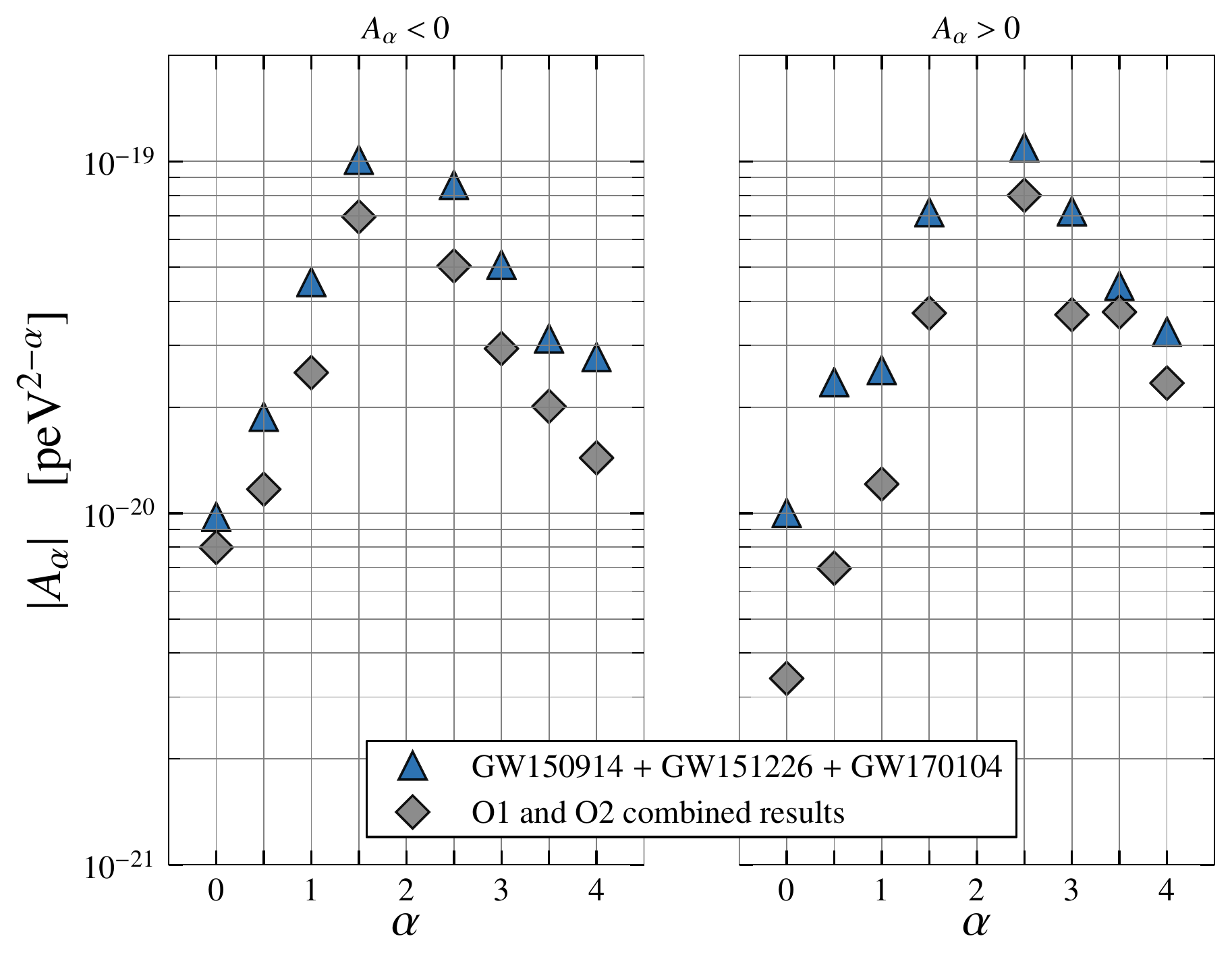}
\qquad
\includegraphics[width=0.48\textwidth]{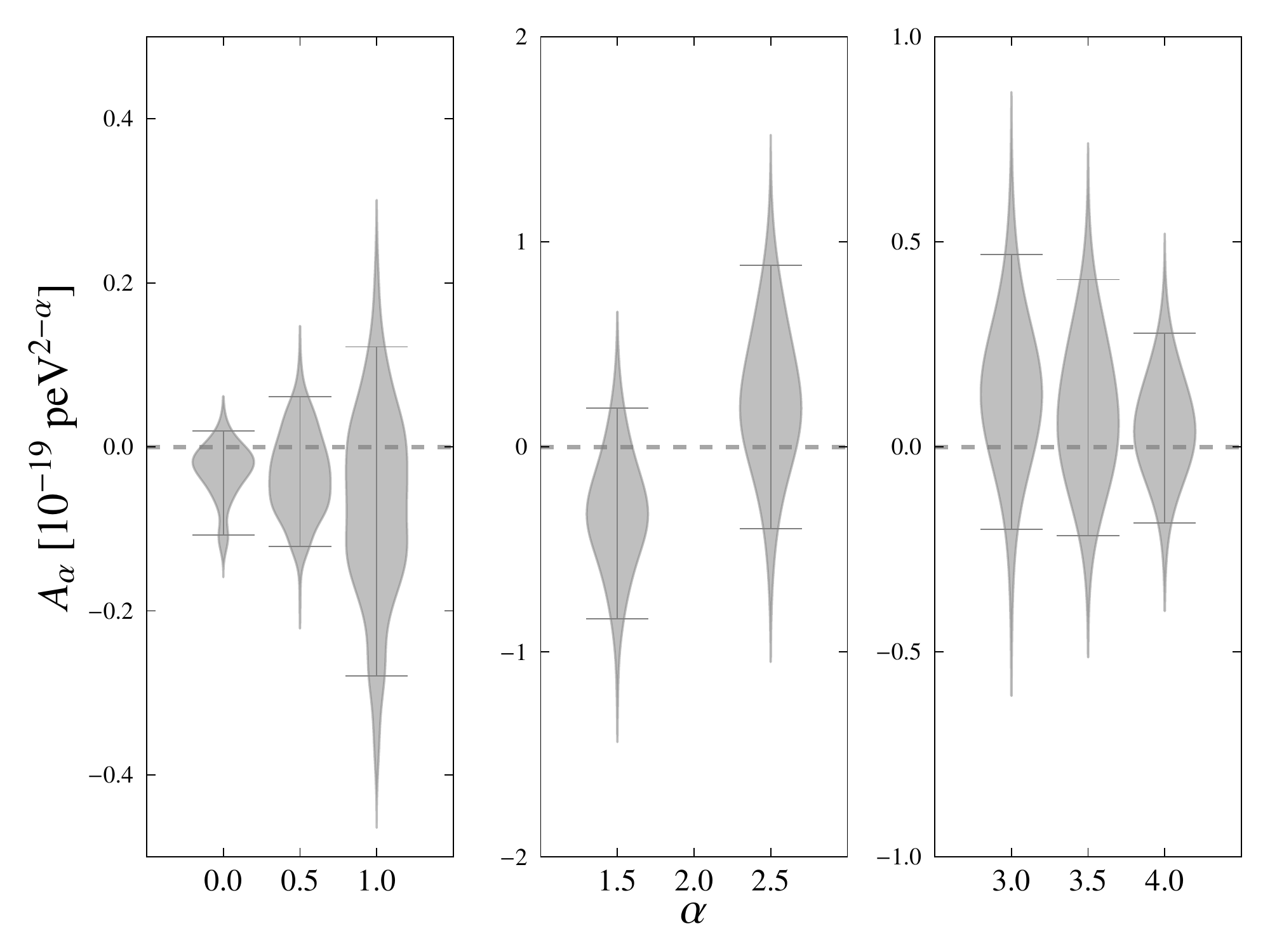}
\caption{\emph{(Left)} 90\% credible upper bounds on the absolute value of the modified dispersion relation parameter $A_\alpha$. We show results for positive and negative values of $A_\alpha$ separately. Specifically, we give the updated versions of the results from combining together GW150914, GW151226, and GW170104, as well as the results from combining together all the events meeting our significance threshold for combined results (see Table~\ref{tab:events}). Picoelectronvolts (peV) provide a convenient scale, because $1 \text{ peV} \simeq h\times 250$~Hz, where $250$~Hz is roughly around the most sensitive frequencies of the LIGO and Virgo instruments. \emph{(Right)} Violin plots of the full posteriors on the modified dispersion relation parameter $A_\alpha$ calculated from the combined events, with the $90\%$ credible interval around the median indicated. Figures reproduced from the main paper.}
\label{fig:prop_results}
\end{figure}

The phenomenological dispersion relation $E^2 = p^2c^2 + A_\alpha p^\alpha c^\alpha$ (where $E$ and $p$ are the energy and momentum of the gravitational waves and $A_\alpha$ and $\alpha \geq 0$ are phenomenological coefficients) can describe several possible extensions of the dispersion relation in GR ($A_\alpha = 0$), including a massive graviton ($\alpha = 0$, $A_0 > 0$) and various Lorentz-violating phenomenologies~\cite{Mirshekari:2011yq,Yunes:2016jcc}. This dispersion relation leads to dispersive propagation of the waves (and thus an observable phase shift) in all cases with $A_\alpha \neq 0$ except for $\alpha = 2$, where it just rescales the speed of propagation with no dispersion. As discussed in the main paper, it is reasonable to take the waveform that would be observed near the source to be given by the GR waveform to good accuracy, and thus the only modifications to the waveform are those from the propagation governed by the modified dispersion relation.

We present the results in Fig.~\ref{fig:prop_results}, finding no indication for dispersive propagation. The individual constraints improve previous results (from the GW170104 paper~\cite{GW170104}) by up to a factor of $\MaximumImprovementFromPreviousBoundsOTwoTGR$.\footnote{The results in the GW170104 paper~\cite{GW170104} were affected by a slight normalization issue, and also had insufficiently fine binning in the computation of the upper bounds. However, we find improvements of up to a factor of $3.4$ when comparing to the combined GW150914 + GW151226 + GW170104 bounds we compute here.} In particular, we improve our previous constraint on the graviton mass by a factor of $\NinetyPercentCombinedGravitonMassBoundOTwoTGRImprovementFromGWOneSevenZeroOneZeroFourPaper$, to $m_g \leq \NinetyPercentCombinedGravitonMassBoundScaledOTwoTGR \times 10^{-23} \text{ eV}/c^2$ ($90\%$ credible level), slightly better than the complementary Solar System constraint using the Yukawa potential of $m_g \leq 6.76  \times 10^{-23} \text{ eV}/c^2$ (90\% confidence level)~\cite{Bernus:2019rgl}. The constraints on modified dispersion obtained from GW170817~\cite{bns-tgr} are not as stringent as the ones from binary black holes, since the source of GW170817 is much closer.

\section{Constraints on alternative polarizations}

For sources observed with three detectors, it is possible to distinguish signals with purely tensor polarizations from those with purely vector or purely scalar polarizations, due to the detectors' different antenna pattern for the different polarizations~\cite{Isi:2017fbj}. One computes the Bayes factors comparing purely tensor polarizations to purely vector or purely scalar polarizations and finds that purely tensor polarizations is indeed preferred, as predicted by GR. By far the best constraints (Bayes factors $> 10^{20}$ in favor of purely tensor polarizations) come from GW170817, with its precise sky localization from the optical counterpart~\cite{bns-tgr}. However, the binary black hole signals still clearly favor purely tensor polarizations, with GW170814 and GW170818 giving Bayes factors of a few tens and hundreds in favor of purely tensor polarizations versus purely vector or scalar polarizations, respectively. The other binary black holes detected with all three detectors do not have a high enough SNR and/or good enough sky localization to provide relevant information.

\section*{Acknowledgments}

The authors gratefully acknowledge the support of the United States
National Science Foundation (NSF) for the construction and operation of the
LIGO Laboratory and Advanced LIGO as well as the Science and Technology Facilities Council (STFC) of the
United Kingdom, the Max-Planck-Society (MPS), and the State of
Niedersachsen/Germany for support of the construction of Advanced LIGO 
and construction and operation of the GEO600 detector. 
Additional support for Advanced LIGO was provided by the Australian Research Council.
The authors gratefully acknowledge the Italian Istituto Nazionale di Fisica Nucleare (INFN),  
the French Centre National de la Recherche Scientifique (CNRS) and
the Foundation for Fundamental Research on Matter supported by the Netherlands Organisation for Scientific Research, 
for the construction and operation of the Virgo detector
and the creation and support  of the EGO consortium. 
The authors also gratefully acknowledge research support from these agencies as well as by 
the Council of Scientific and Industrial Research of India, 
the Department of Science and Technology, India,
the Science \& Engineering Research Board (SERB), India,
the Ministry of Human Resource Development, India,
the Spanish  Agencia Estatal de Investigaci\'on,
the Vicepresid\`encia i Conselleria d'Innovaci\'o, Recerca i Turisme and the Conselleria d'Educaci\'o i Universitat del Govern de les Illes Balears,
the Conselleria d'Educaci\'o, Investigaci\'o, Cultura i Esport de la Generalitat Valenciana,
the National Science Centre of Poland,
the Swiss National Science Foundation (SNSF),
the Russian Foundation for Basic Research, 
the Russian Science Foundation,
the European Commission,
the European Regional Development Funds (ERDF),
the Royal Society, 
the Scottish Funding Council, 
the Scottish Universities Physics Alliance, 
the Hungarian Scientific Research Fund (OTKA),
the Lyon Institute of Origins (LIO),
the Paris \^{I}le-de-France Region, 
the National Research, Development and Innovation Office Hungary (NKFIH), 
the National Research Foundation of Korea,
Industry Canada and the Province of Ontario through the Ministry of Economic Development and Innovation, 
the Natural Science and Engineering Research Council Canada,
the Canadian Institute for Advanced Research,
the Brazilian Ministry of Science, Technology, Innovations, and Communications,
the International Center for Theoretical Physics South American Institute for Fundamental Research (ICTP-SAIFR), 
the Research Grants Council of Hong Kong,
the National Natural Science Foundation of China (NSFC),
the Leverhulme Trust, 
the Research Corporation, 
the Ministry of Science and Technology (MOST), Taiwan
and
the Kavli Foundation.
The authors gratefully acknowledge the support of the NSF, STFC, MPS, INFN, CNRS and the
State of Niedersachsen/Germany for provision of computational resources.
This document is LIGO-P1900146-v5.

\section*{References}

\bibliography{cbc-group_Moriond}

\end{document}